%% file: paper.tex
\documentclass[sigconf,preprint]{acmart}

\usepackage{tikz}
\usetikzlibrary{calc,decorations.pathreplacing}
\usepackage{varwidth}
\usepackage{color}
\usepackage{xcolor}
\usepackage{amsthm}
\usepackage{mathcomp}
\usepackage{listings}
\usepackage{enumitem}

\usepackage{makecell}
\usepackage{algorithm}
\usepackage{algpseudocode}

\renewcommand\footnotetextcopyrightpermission[1]{} 
\begin{document}



\begin{abstract}
    Message logging protocols are enablers of local rollback, a more efficient alternative to global rollback, for fault tolerant MPI applications.
    Until now, message logging MPI implementations have incurred the overheads of a redesign and redeployment of an MPI library, as well as continued performance penalties across various kernels.
    Successful research efforts for message logging implementations do exist, but not a single one of them can be easily deployed today by more than a few experts.
    In contrast, in this work we build efficient message logging capabilities on top of an MPI library with no message logging capabilities; we do so for two different HPC kernels, one with a global exchange pattern (CG), and one with a neighbourhood exchange pattern (LULESH).
    While our library of choice ULFM detects failure and recovers MPI communicators, we build on that to then restore the intra- and inter-process data consistency of both applications.
    This task turns out to be challenging, and we present the methodology for doing so in this work. 
    In the end, we achieve message logging capabilities for each kernel, without the need for an actual message logging runtime underneath.
   On the performance side, we match state-of-the-art solutions and (a) eliminate event logging and the event logger component altogether, and (b) design a hybrid protocol, which gracefully shifts between global and local rollback, depending on the available payload logging memory.
    Such a hybrid protocol between local and global rollback has not been previously proposed to our knowledge.
    Our extensions span a few hundred lines of code for each kernel, are open-sourced, and enable local and global rollback after process failure.
\end{abstract}

\title{Implementing Efficient Message Logging Protocols as MPI Application Extensions}

\author{Kiril Dichev}
\affiliation{
\institution{Queen's University Belfast\\Belfast, United Kingdom}
}
\email{K.Dichev@qub.ac.uk}

\author{Dimitrios~S. Nikolopoulos}
\affiliation{
\institution{Queen's University Belfast\\Belfast, United Kingdom}
}
\email{D.Nikolopoulos@qub.ac.uk}

\maketitle
 
\input{intro}
 
 \input{related-work}

\input{transactions}

\input{event-logging}

\input{inter-process-consistency}

\input{send-wrapper}
\input{capping}

\input{cg-outline}

 \input{experiments}
 \input{conclusion}
 
 \bibliographystyle{plain}
\bibliography{references}

\end{document}

%% file: intro.tex

 \section{Introduction}

It is widely accepted that compute clusters and supercomputers are transitioning towards systems of millions of compute units to satisfy the requirements of compute-intensive parallel scientific applications.
With this increase in compute components, a proportional decrease in the Mean-Time-Between-Failure (MTBF) across parallel executions will follow \cite{Schroeder2010,Zheng2012}, which would make highly scalable parallel application runs infeasible without integrating resilience.

In this manuscript, we focus on recovery from fail-stop errors, i.e. any failures leading to the unexpected termination of an MPI process and the loss of its data; a node crash is among the possible causes of fail-stop errors.
For such failures, checkpoint/restart (C/R) strategies are commonly used; they introduce time redundancy due to the rollback of execution but require fewer additional resources than resource replication techniques.
The recent advances in fault-tolerant MPI library implementations, such as ULFM~\cite{Bland2013}, have integrated efficient and scalable detection and recovery primitives to allow MPI applications to deal with failures.
C/R with global rollback is the most widely used fault tolerance technique in HPC applications today; in global rollback, all processes roll back to the last globally consistent checkpoint, and continue execution.
Global rollback can be universally applied to all types of application kernels, and its significant advantages in failure-prone executions are widely known and accepted. 

In order to objectively argue which aspect of recovery should be optimised, quantitative data should be analysed. 
A large-scale study of the highly optimised S3D application on the Titan supercomputer from Gamell et al.~\cite{Gamell2014} shows that the rollback phase is the main culprit.
Therefore, optimisations reducing the amount of rollback need to be studied in the area of fault tolerance research.

There are multiple ways to optimise rollback after failure.
One possibility, which has seen considerable attention in recent years, is the classic direction of fault tolerance studies -- checkpoint/restart. 
Optimised checkpoint/restarts naturally lead to reduced rollback, even if all processes need to roll back.

An entirely orthogonal idea is to allow surviving processes to perform as little recompute as possible, by only rolling back failed processes after failure.
This is the realm of local rollback recovery, and of message logging protocols, the enablers of local rollback.
Upon failure, message logs can be replayed to a failed process, and if correctly designed, message logging can allow survivors to not roll back, a potential advantage over all traditional global rollback strategies.
Various message logging protocols \cite{Alvisi1998}, exist, such as pessimistic, optimistic, and causal message logging.
A large body of research exists, particularly with the MPICH-V and MPICH-V2 projects, which studied various runtime designs and protocols for message logging.
Unfortunately, various kernels consistently show large runtime overheads even for well-designed message logging runtimes.
In addition, both of these projects have disappeared as research efforts.
More recent MPI implementations also struggle to provide easy to deploy message logging support; e.g. Charm++ has abandoned such support in their latest stable releases.
The most up-to-date message logging protocol can be found in the pessimistic VProtocol within the Open MPI codebase; however, the support is only partial (payload logging), and just a stepping stone to a complete message logging runtime, and needs additional features to be made functional, not all of which are freely available and easy to use~\cite{Losada2019}.
Today, not a single easy-to-use MPI implementation with message logging capabilities exists.

In this work, we challenge this landscape and implement a message logging protocol as an efficient and configurable extension of HPC application kernels.
We describe the various issues we faced and how we resolved them.
We rely on ULFM, the de-facto fault-tolerant MPI implementation, for capabilities such as failure detection and communicator recovery.
No message logging MPI library, or other components, such as external payload or event loggers, are required.
Our HPC kernels of choice are the MPI version of the NAS CG benchmark, and the larger proxy app LULESH from LLNL.
The former has always been among the most challenging kernels due to its global and intensive communication, whereas the latter represents a different class of stencil-like codes with neighbourhood exchange.

After implementing message logging as an extension to applications, we also achieve some interesting optimisations, such as (a) no event logging or event loggers for these kernels (b) configurable payload logging, since payloads can be conceived as fine-grained application checkpoints.

The novel contributions of this work are:
\begin{itemize}
    \item A methodology for implementing message logging protocols as compact extensions of HPC kernels, exemplified for two HPC kernels. The methodology includes:
    \begin{itemize}
    \item The concept of iterations as transactions, which helps us resolve intra-process data consistency issues, as well as eliminate event logging
    \item The snapshot of the computational wave after failure, which can be used to replay messages and restore inter-process data consistency, on top of the intra-process consistency.
    \end{itemize}
    \item Due to the scope of payload logging within the application, a hybrid scheme with graceful degradation between local and global rollback is introduced, depending on available payload capacity.
\end{itemize}

The implemented codes are open-sourced, and rely on other open-sourced codes, such as NPB, LULESH, and ULFM.

The paper is organised as follows: the related work in message logging is presented in Sect.~\ref{sec:related}. In Sect.~\ref{sec:intra-proc}, we introduce the concept of transactional MPI iterations, which allows us to restore intra-process data consistency.
We then focus on inter-process data consistency in Sect.~\ref{sec:inter-process}.
We continue with the technical aspects, such as the send wrapper (Sect.~\ref{sec:send-wrapper}) and replay routine (Sect.~\ref{sec:capping}) implementations.
We briefly summarise the extensions required for CG and LULESH in Sect.~\ref{sec:loc}.
Our experiments are shown in Sect.~\ref{sec:experiments}, after which we conclude the paper (Sect.~\ref{sec:conclusion}).

%% file: related-work.tex

 \section{Related Work}
 \label{sec:related}
The idea of implementing message logging protocols in application kernels is not entirely new.
Gamell et al.~\cite{Gamell2017} outlined the use of sender-based logs for a stencil code called S3D, integrating the rollback on top of ULFM; while the authors never open-source their work, we clearly outline the design we use for two kernels.
We suspect that LULESH has some similarities in its neighbourhood exchange patterns to S3D; However, the NAS CG kernel is more challenging with its global communication patterns, such as the allreduce, and possibly better placed to stress message logging implementations.

Within MPI runtimes, message logging protocols are a widely studied area of fault tolerance.
All of them are generally applicable to various kernels, whereas our contributions are applied per HPC application as kernel extensions.
The seminal paper of Strom~\cite{Strom1985} first introduced optimistic message logging protocols; it also formalised orphans and determinants.
Researchers have proposed an optimised version of optimistic message logging~\cite{Ropars2006,Bouteiller2009}, which reduces the extent of piggybacked determinants, due to the deterministic nature of communication events.
Implementations of causal message logging protocols in MPI libraries include MPICH-V~\cite{Bosilca2002}, MPICH-V2~\cite{Bouteiller2003mpich}, and Charm++~\cite{Meneses2011}.
The work of Losada et al.~\cite{Losada2019} implements the most recent variation on pessimistic message logging protocols; it employs the VProtocol in Open MPI \footnote{not freely available at the time of writing}, and a source-to-source compiler called CPPC.
Our work, in contrast to pessimistic, optimistic, and causal message logging protocols, entirely disables event logging; we statically analyse the underlying applications, and guarantee the absence of non-deterministic events that need to be replayed.
In this respect, the only contribution that performs a similar optimisation is of Bouteiller et al.~\cite{Bouteiller2010}.
Their pessimistic protocol is highly optimised to reduce event logging to challenging non-deterministic events; for example, any-sender non-blocking receives are such events.
The authors log no events for the NAS CG benchmark, similarly to our work, even though using a runtime mechanism, rather than a kernel extension.
 
In terms of exploring determinism of HPC codes, interesting contributions exist, which similarly to our work look for optimisations based on the absence of non-determinism.
Guermouche et al.~\cite{Guermouche2011} also carefully examines the communication properties of various kernels, and argue that most codes are send-deterministic, which allows for more efficient protocols to be implemented; the focus is specifically on causal message logging protocols.
In a rather complex runtime extension, the authors manage to reduce the number of logged messages, while still preserving the causality of their message logging algorithm, for send-deterministic applications.
 However, to reduce the message logs, the protocol introduces some overhead to messages, such as piggybacked data (e.g. epoch numbers); often, a large set of processes needs to roll back, while in our proposal only the failed process rolls back.
There are some similarities to our optimisations in these related works; we similarly perform ``static analysis of applications consisting of looking at each communication pattern and studying its deterministic nature''.
However, there are numerous differences in the underlying techniques: among others, we work in the context of the application, and not the MPI implementation.

In terms of payload logging, and optimisations for that, Losada et al. cap the message logs by observing the checkpoint intervals, and discard older logs.
They enable this optimisation via application instrumentation with a source-to-source compiler (CPPC).
We also cap the payload logs to be contained within a checkpoint interval.
However, we do not stop there -- the capping of payload logging may continue gradually, allowing for a graceful degradation between global and local rollback.
To the best of our knowledge, no message logging protocol has ever implemented such a hybrid solution.

We also give some background to the use of the NAS CG benchmark as a challenging problem for message logging protocols here.
 In a study for MPICH-V, Bouteiller et al.~\cite{Bouteiller2005} examine the NAS benchmarks, CG, BT, LU, and FT; the authors find that LU and CG (in that order) cause the most memory and compute overheads for an implementation of a causal message logging protocol.
The same researchers~\cite{Bouteiller2010} later introduce an optimised pessimistic message logging protocol, which among others reduces overhead for kernels with non-deterministic patterns, such as LU.
 After optimisation, they observe the highest runtime overhead with NAS CG, at 5\%, for 64 node runs.
 More recently, Meneses et al.~\cite{Meneses2011}  study the benchmark CG, MG, BT, and DT from the same suite, with implementation of pessimistic and causal message logging protocols, and find that pessimistic message logging CG shows 17\% runtime overheads, but that a causal message logging shows significant improvements, with around 1\% runtime overheads for the CG kernel.

 As a summary and comparison with closely related work, we provide Table  \ref{tab:comparison-table}; whenever possible, we provide some reference numbers for CG; such numbers do not exist in these contributions for LULESH.
  
\input{comparison-table}

In terms of availability, MPICH-V and MPICH-V2 are deprecated and not available, and the message logging capabilities in Charm++ are not functional in recent stable releases.
The only currently usable message logging protocol, the pessimist Vprotocol in Open MPI, is in an experimental stage, and not ready for use out-of-the-box; recently proposed optimised protocols are not currently available for testing~\cite{Losada2019}. 
We believe this landscape clearly motivates our work to implement message logging capabilities as application extensions on top of popular and widely-used MPI implementations.
In this way, compact implementations of message logging as kernel extensions can be tested or even ported from our open-sourced work.

%% file: comparison-table.tex

\begin{table}
\begin{small}
\begin{tabular}{|p{2.5cm}|p{1cm}|p{1cm}|p{1cm}|p{1cm}|}
 & \cite{Bouteiller2010} & \cite{Meneses2011} & \cite{Bouteiller2009} &this work\\
\hline
protocol & pessi\-mistic & causal & opti\-mistic & event-less \\
\hline
implemented in & Open MPI & Charm++ & Open MPI & CG and LULESH (extended)\\
\hline
event logging required & not for CG & yes & yes & no \\
\hline
message logging required & yes & yes & yes & yes \\
\hline
requires tracking determinants & no & yes & yes & no \\
\hline
optimised determinant protocol? & -- & yes & yes & -- \\
\hline
applicability & general & general & general & per kernel \\
\hline
overheads & $\approx 5\%$ for CG & $\approx 1\% $ for CG & $\approx 5\%$ for CG  & fully configurable \\
\hline
main culprit & payloads &  not detailed & payloads & payloads \\

\hline
\end{tabular}
\end{small}
\caption{A table of related work exploring message logging. We also compare the related work to ours in various important aspects, including NAS CG results where available.}
\label{tab:comparison-table}
\end{table}

%% file: transactions.tex
\section{Intra-Process Data Consistency}
\label{sec:intra-proc}

In this section, we guarantee data consistency within a process, even after an iteration is rolled back; we deal with inter-process data consistency in a later section.

\subsection{Transactional MPI Iterations}
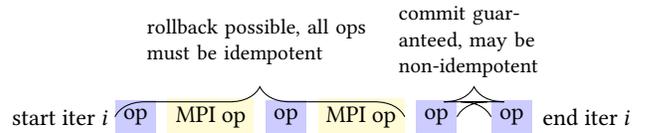
\begin{figure}
\begin{tikzpicture}[node distance = 1cm]
\node[rectangle] (start) {start iter $i$};
\node[fill=blue!20,fill,rectangle,right of=start] (a) {op};
\node[fill=yellow!20,rectangle,right of=a] (b) {MPI op};
\node[fill=blue!20,rectangle,right of=b] (c) {op};
\node[fill=yellow!20,rectangle,right of=c] (d) {MPI op};
\node[fill=blue!20,rectangle,right of=d] (e) {op};
\node[fill=blue!20,rectangle,right of=e] (f) {op};
\node[rectangle,right of=f] (end) {end iter $i$};
\draw [decorate,decoration={brace,amplitude=0.4cm},yshift=1cm] (a.west) -- (d.east) node [font=\small,midway,yshift=1cm,text width=3cm] {rollback possible, all ops must be idempotent};
\draw [decorate,decoration={brace,amplitude=0.4cm},yshift=1cm] (e) -- (f) node [font=\small,midway,yshift=1cm,text width=2cm] {commit guaranteed, may be non-idempotent};
\end{tikzpicture}
\caption{Illustration of ensuring the transactional nature of an iteration $i$: until the last MPI op in iteration $i$, a rollback is possible, therefore operations need to be idempotent. After the last MPI op call within an iteration, a commit is guaranteed, and operations may change global data irreversibly.}
\label{fig:fail-safe-iteration}
\end{figure}

To guarantee data consistency of each process after failure, we take the view that an iteration can be designed as an atomic transaction.
The fundamental property of a transaction is that it can either be committed or aborted; if aborted, it leaves the application in its pre-transaction state.
Like any MPI code, each iteration of an HPC kernel consists of two types of operation -- local operations, or non-local operations (MPI calls).
For an iteration to have transactional nature, we must be able to roll it back whenever a failure is detected.
In fault-tolerant MPI implementations, only in MPI functions can report the failure of another process.
In effect, this reduces the problem to the illustration of Fig. \ref{fig:fail-safe-iteration}.
All phases including the very last MPI operations in an iteration (in yellow) may be repeated.
Therefore, all operations need to be idempotent.
If, in contrast, some operation is non-idempotent, the rollback cannot happen without further action.
For example, if one of the local operations (in blue) preceding the last MPI operation is an increment on global data, the iteration is not transactional in nature.
A repetition of iteration $i$ will lead to incorrect global data, since the increment of global data would be performed twice.
A kernel iteration is transactional in nature if it performs all non-idempotent operations only after the very last MPI call in an iteration.

We hold that all non-transactional iterations can be made transactional, with some memory duplication, and in some cases added copy operations.
For example, the illustrated code in Alg.~\ref{fig:non-trans-iterations} is non-transactional: an increment on global data is a non-idempotent operation, and it happens before an MPI call.
If we perform a rollback of this iteration, not reading the last checkpoint, this may result in $x$ being incremented twice for the same iteration.
The solution to this problem is outlined in Alg.~\ref{fig:trans-iterations}, and consists of guaranteeing that no non-idempotent operations are called \textit{until after the last MPI call} in the iteration.
If a failure is detected during any MPI call, the transaction (or iteration) can be rolled back with no further action, and without undesired side effects on global data, since all operations are idempotent.
After this call, the occurrence of a failure cannot be detected in the same iteration anymore, i.e. the iteration would be ``committed''.

\begin{algorithm}
\begin{lstlisting}[language=c,basicstyle=\small]
// non-transactional
for (int i=0; i<NITER; i++) {
	x++; // non-idempotent, re-run bad!
	MPI_Sendrecv(&x,...,&y,...);
	// further local operations
	...
}
\end{lstlisting}
\caption{Illustration of a non-transactional MPI iteration}
\label{fig:non-trans-iterations}
\end{algorithm}

\begin{algorithm}
\begin{lstlisting}[language=c,basicstyle=\small]
// transactional
for (int i=0; i<NITER; i++) {
	x_tmp = x;
	x_tmp++;
	MPI_Sendrecv(&x_tmp,...,&y_tmp,...);
	// non-idempotent, but will never be reversed
	x = x_tmp; 
	y = y_tmp;
	// further local operations
	...
}
\end{lstlisting}
\caption{Illustration of a transactional MPI iteration}
\label{fig:trans-iterations}
\end{algorithm}


\subsection{CG and LULESH iterations as transactions}

\input{cg-phases}

Consider the local operations and non-local operations illustrated in Fig. \ref{cg:phases} for the NAS CG and LULESH benchmarks.
For each kernel, local operations are outlined in blue and brown, and non-local MPI calls, are outlined in yellow.
In their original implementations, both CG and LULESH are not transactional in nature; they write into global data before their last MPI operations in each iteration, that is both kernels have non-idempotent operations which means that any rollback will break the data consistency of the application, unless code modifications similar to these shown in Alg.~\ref{fig:trans-iterations} are made.
We did this for both kernels, and these modifications are highlighted in brown in the figure.
The recovery of failed processes includes a sequence of stages, highlighted as red boxes in the figure; some of them must be performed by a fault-tolerant MPI implementation (such as ULFM).
Others include our extensions to restore data consistency, and to provide local rollback.

\begin{itemize}
\item the MPI implementation must have a failure detection mechanism. The failure detection can only be triggered during MPI calls, when a surviving MPI process communicates with either a failed process, or a process which has already revoked the communicator.
\item the developer can register an error handler routine to handle failures. The routine generally has 2 stages: \textbf{(a)} recovery of the MPI communicator (MPI-specific) and \textbf{(b)} recovery of the application state (application-specific).
\end{itemize}

Our work is concerned with restoring data consistency within a process, followed by global data consistency between MPI processes as well.
We note here that we strictly require an MPI implementation to provide primitives for failure detection and communicator recovery; without these primitives, our contribution would fail.
This requirement is rather generic, and is true also for automated global rollback; MPI communicator recovery is essential if any rollback is dynamic rather than triggered externally.
Our solution is embedded in the application, and no message logging capabilities are required of the MPI runtime, while local rollback is still achieved.

It is worth noting that transactional logic of iterations was initially conceived for practical reasons; upon process failure, an error handler interrupts the control flow of the surviving processes.
After the successful MPI communicator recovery, we need to pass control back to the application kernel -- but where?
If we interpret an iteration as a transaction, our task is easy -- we roll back to the start of the current iteration.
Note that this control flow after failure differs from its counterpart with message logging protocols in runtimes.
For example, pessimistic protocols synchronously log events before a message is passed, so no rollback of survivors to earlier program stages is required, or even possible.

A failed process is always restarted from the last global checkpoint.

%% file: cg-phases.tex

\begin{figure}
\begin{tikzpicture}[style = {font=\tiny}, node distance=0.6cm]
\node (A1) [fill=black!25,fill,draw,rotate=90] {begin iteration $i$};
\node (T1) [right of=A1,fill=brown!25,text width=2.5cm,rotate=90] {deep copy $z,r$ to temp buffer};
\node (A) [right of=T1,fill=blue!25,text width=3cm,rotate=90] {local $q = A.p$ compute};
\node (B) [right of=A,fill=yellow!25,text width=3cm,rotate=90]{send $w$ / recv $q$ between 2 reduce partners sequentially, with $w$ updates};
\node (C) [right of=B, fill=yellow!25,text width=3cm,rotate=90]{send $w$ / recv $q$ between transpose partners};
\node (D) [right of=C, fill=blue!25,text width=3cm,rotate=90] {clear $w$; local compute $p.q$};
\node (E) [right of=D, fill=yellow!25,text width=3cm,rotate=90] {send $sum$  - receive $d$ between reduce partners sequentially, performing local updates after each send};
\node (F) [right of=E, fill=blue!25,text width=3cm,rotate=90] {local compute of $\alpha, z, r$};
\node (G)  [right of=F, fill=yellow!25,text width=3cm,rotate=90] {send $sum$ -recv $rho$ between reduce partners sequentially, with $sum$ updates};
\node (T2) [right of=G,fill=brown!25,text width=2.5cm,rotate=90] {deep copy temp buffer to $z,r$};
\node (H)  [right of=T2, fill=blue!25,text width=3cm,rotate=90] {local updates of $\beta, p$};
\node (I) [right of=H, fill=black!25,fill,draw,rotate=90] {end iteration $i$};
\node[fill=red!25,draw] (handler) at ($(A1)!0.5!(I) + (0,2)$) {error handler};
\node[fill=red!25,draw,left of=handler,text width=1cm,node distance=2cm] (commrec) {ULFM comm recovery};
\node[fill=red!25,draw,left of=commrec,text width=1cm,node distance=1.5cm] (replay) {replay routine};
\path[draw,red,->] (B.east) --  (handler);
\path[draw,red,->] (E.east) --  (handler);
\path[draw,red,->] (C.east) --  (handler);
\path[draw,red,->] (G.east) --  (handler);
 \path[draw,red,->] (handler.west) -- (commrec.east);
 \path[draw,red,->] (commrec.west) -- (replay.east);
 \path[draw,red,->] (replay.south) -- (A1.north);
 
\draw [decorate,decoration={brace,amplitude=0.4cm,raise=2.2cm}] (A1.north) -- (I.north) node [font=\small,midway,yshift=2.7cm,text width=3cm] {NAS CG iteration};
 
  \begin{scope}[shift={(0,-5)}]
  \node (A1) [fill=black!25,fill,draw,rotate=90] {begin iteration $i$};
  \node (T1) [right of=A1,fill=brown!25,text width=2cm,rotate=90] {deep copy domain to temp buffer};
\node (A) [right of=T1,fill=yellow!25,text width=3cm,rotate=90] {LagrangeNodal (sends/recvs)};
\node (B) [right of=A,fill=yellow!25,text width=3cm,rotate=90] {LagrangeNodal (sends/recvs)};
\node (C) [right of=B,fill=blue!25,text width=3cm,rotate=90]{LagrangeNodal};
\node (D) [right of=C, fill=yellow!25,text width=3cm,rotate=90]{LagrangeElements (sends/recvs)};
\node (E) [right of=D, fill=blue!25,text width=3cm,rotate=90] {LagrangeElements};
\node (F) [right of=E, fill=yellow!25,text width=3cm,rotate=90] {sends/recvs};
\node (G) [right of=F, fill=blue!25,text width=3cm,rotate=90] {CalcTimeConstraints};
  \node (T2) [right of=G,fill=brown!25,text width=2cm,rotate=90] {deep copy temp buffer to domain};
\node (H) [right of=T2, fill=black!25,fill,draw,rotate=90] {end iteration $i$};
\node[fill=red!25,draw] (handler) at ($(A1)!0.5!(H) + (0,2)$) {error handler};
\node[fill=red!25,draw,left of=handler,text width=1cm,node distance=2cm] (commrec) {ULFM comm recovery};
\node[fill=red!25,draw,left of=commrec,text width=1cm,node distance=1.5cm] (replay) {replay routine};
\path[draw,red,->] (A.east) --  (handler);
\path[draw,red,->] (B.east) --  (handler);
\path[draw,red,->] (D.east) --  (handler);
\path[draw,red,->] (F.east) --  (handler);
 \path[draw,red,->] (handler.west) -- (commrec.east);
 \path[draw,red,->] (commrec.west) -- (replay.east);
 \path[draw,red,->] (replay.south) -- (A1.north);
 
 \draw [decorate,decoration={brace,amplitude=0.4cm,raise=2.2cm}] (A1.north) -- (H.north) node [font=\small,midway,yshift=2.7cm,text width=3cm] {LULESH iteration};
 \end{scope}
 
\end{tikzpicture}
\caption{Different phases of NAS CG and LULESH iterations. Phases in blue and brown are local updates. Phases in yellow are MPI communication events, and detectors of failure. We implement the copy to/from temp buffers (brown) as enablers of transactional iterations. Phases in red provide recovery of communicators (ULFM), as well as local rollback via our replay routine.}
\label{cg:phases}
\end{figure}
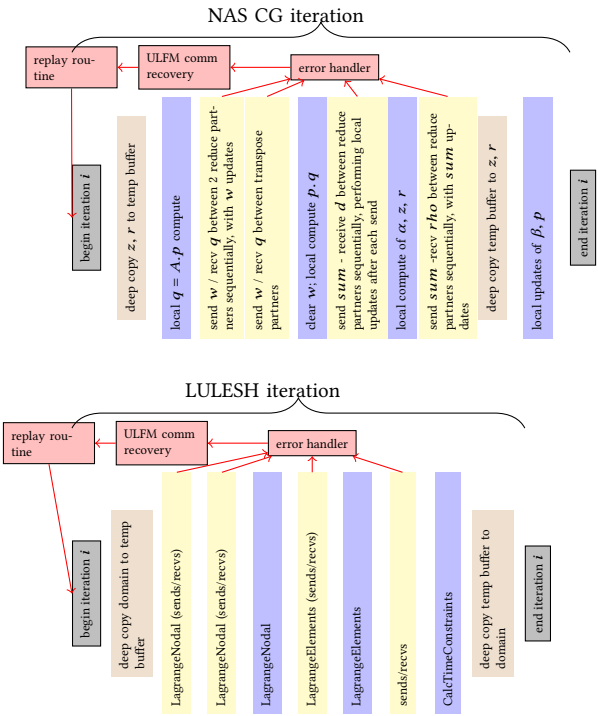

%% file: event-logging.tex

\subsection{Why we need no event logging/logger}

Event logging has always been an integral part of any message logging protocol within MPI.
In the context of MPI applications, every send and receive is usually logged, since the runtime has no concept of the application state, and needs to recover each communication in case of a failure.
Event logging usually incurs an overhead proportional to the latency on the underlying network, since event loggers are quite often placed on different nodes from the application processes nodes, and the popular family of pessimistic message logging protocols requires every event to be synchronously logged before progress.
There have been some optimisations in event logging in the past.
For example, optimistic message logging has previously removed the requirement of synchronously logging events, masking some of the incurred overheads.
Another interesting optimisation~\cite{Bouteiller2010} is the ability of an optimised implementation to only log challenging events (such as any-sender receives), which in effect result in similar disabling of event logging as our work for some kernels, e.g. the NAS CG benchmark.


In this work, we take a substantially different view to events.
Since we implement our protocol as extension of application kernels, it is not MPI communication events that are key, but application events that irreversibly modify the data of a process.
As a consequence of this viewpoint, we can try to reduce the number of these irreversible events.
This view in itself does not eliminate event logging.
Consider, for example, the outline of the different phases of the LULESH mini-app of Fig. \ref{cg:phases}.
Every yellow phase entails numerous non-blocking send/recv calls (exchanges with neighbours in three dimensions), which normally would need to be replayed in the same order in a log-enabled recovery.
However, by conceiving each iteration as a \textit{transaction}, we can abort entire iterations, and eliminate event logging altogether.
During recovery each application only needs to provide its current iteration.
The simplification is significant; since each process can safely ignore the exact phase within its iteration where a failure was detected.

As a consequence of disabling event logging or event loggers, we do not carry the runtime overheads incurred in MPI implemenations.
In contrast, runtime-based logging employs event loggers as remote processes, including long-existing efforts such as MPICH-V, or newer implementations such as Open MPI's VProtocol.
We still need payload logging, as an extension of each MPI process. 

We note that our design is currently applied with some assumptions on the underlying kernel in mind:
\begin{itemize}
\item a kernel is deterministic (but not necessarily send-deterministic), which includes every HPC kernel we are aware of (see \cite{Guermouche2011} for overview)
\item a coordinated checkpointing scheme, possibly asynchronous, is in place, for the given kernel.
\end{itemize}

The fact that we require an explicitly or implicitly coordinated checkpointing scheme carries important simplifications for us, e.g. the fact that orphan or in-transit messages can never occur.

%% file: inter-process-consistency.tex
\section{Inter-Process Data Consistency after Failure}
\label{sec:inter-process}

Despite the fact that the transactional nature of iterations now guarantees that the data of each process remains consistent after failure, inter-process data consistency is an unresolved issue.
Unfortunately, it is a very likely outcome that MPI processes detect failure in different iterations, and therefore the application will have an inconsistent global data state.

Following scenarios are all likely, even in tightly-coupled HPC codes:
\begin{itemize}
\item Survivors are very likely to be in different iterations after failure detection.
\item Restarted process and survivors are almost certainly in different iterations upon failure detection.
\end{itemize}

We will detail the issues and their solutions in this section.

 \subsection{Survivors -- different iterations}
 \label{sec:replay-mpi}
An important issue for most MPI kernels, even when they have desirable properties such as determinism, is the non-transactional nature of MPI point-to-point communication calls.
The completion of an MPI point-to-point call, for example, is not a guarantee of a delivered message; the most obvious example is the \verb1MPI_Send1 call, which can be buffered locally (for example, small-size message and the internal use of eager vs rendezvous protocols).
 A matching receive may still be unsuccessful due to a failure of the receiving process, leading to inconsistencies in the global state.
This non-transactional nature of MPI point-to-point communication requires us to carefully consider how to recover (see also e.g. ~\cite{Dichev2018,Losada2019,Laguna2014}).
 
 \input{16proc}

\begin{figure}
\includegraphics[width=0.35\textwidth]{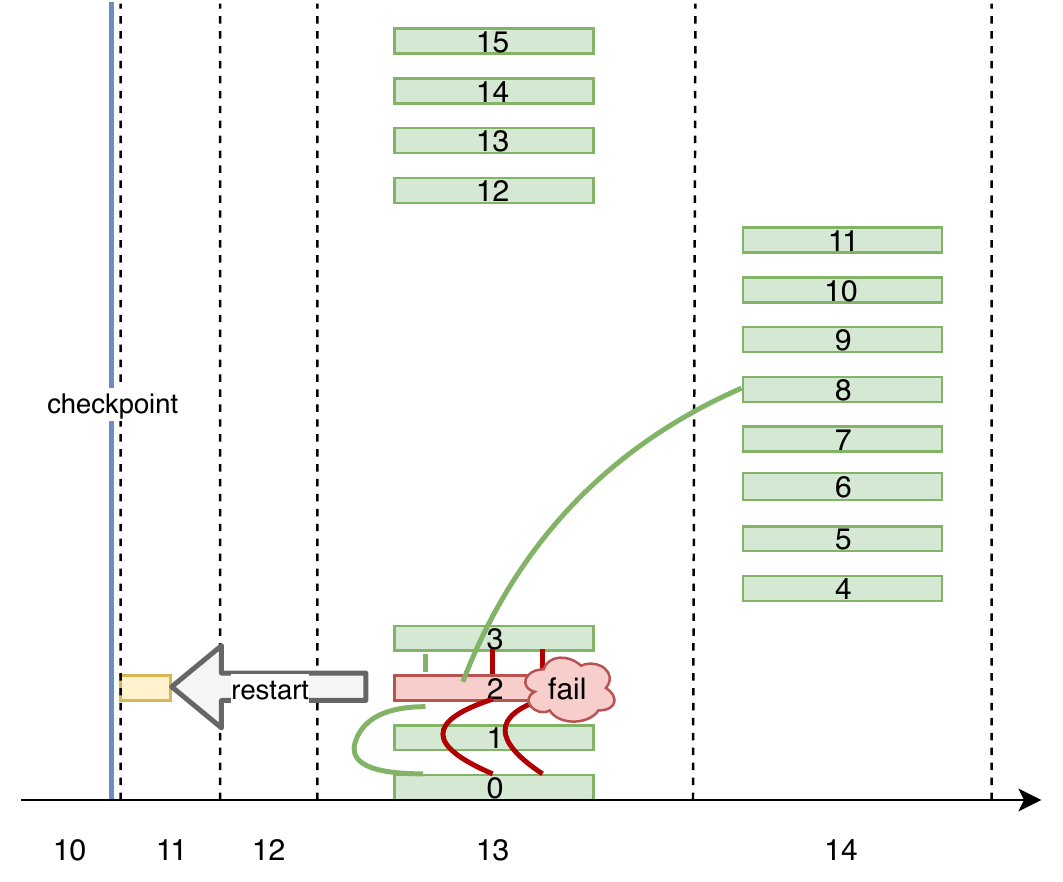}
\caption{Illustration of a snapshot of a real failure scenario for cg.C.16 (16 processes). Process 2 fails in iteration 13, and is restarted at the last checkpoint. Not all survivors detect the failure in the same iteration.}
\label{fig:failure-scenario}
\end{figure}

Consider a 16 process run with CG (cg.C.16), where a square matrix is partitioned among processes as illustrated in Fig. \ref{fig:16procs}.
There are 2 reduce partners (black edges), and one transpose partner (red edges), for each process.

A snapshot of an actual failure scenario for this setting is illustrated in Fig. \ref{fig:failure-scenario}.
All ranks write their part of a global checkpoint every 10 iterations, the most recent checkpoint being at the end of iteration 10.
Rank 2 crashes unexpectedly in the middle of iteration 13, just after the exchange with the transpose partner 8.
Each process can only notice the failure in an MPI call (phases highlighted in yellow for CG and LULESH in Fig. \ref{cg:phases}).
Rank 0 and/or 3 will immediately notice the failure of rank 2, since they cannot possibly receive data they need.
The communicator will be revoked. 
Yet it is not necessary that all MPI processes notice the revoked communicator and trigger recovery in the same iteration.
In a typical scenario, processes \{0,1,3,12,13,14,15\} are notified in the same iteration 13, being direct or ``close'' communication partners with 2 after its failure.
However, processes \{4,\dots,11\} are notified while already in the following iteration 14 -- their completion of iteration 13 was successful, having received their required messages from their communication partners.

Due to the outlined non-transactional nature of MPI P2P calls, a number of surviving processes are in iteration 14 and others in iteration 13. All processes abort their current iteration, and restart it. The iteration-14 survivors are now responsible for replaying the message contents of iteration 13 to all survivors left behind, in order to restore consistency.

\subsection{Survivors and restarted process -- different iterations}
\label{sec:replay-normal}

As illustrated in Fig. \ref{fig:failure-scenario}, process 2 fails at iteration 13, reloads from the last checkpoint it has written, and starts at iteration 11.
The survivors have obviously progressed further, being in iterations 13 and 14.

To restore the consistency, process 2 would read the checkpoint of iteration 10, and iterate through til iteration 13, receiving replayed messages. However, process 2 must not send any messages to any peers til iteration 13, since these messages have already been successfully sent (otherwise survivors would not have progressed).
Of the surviving processes, 0 and 3 are reduction peers of 2, and 8 is the transpose peer of 2. 
Therefore, peers 0, 3, and 8, all need to replay messages from iterations 11, 12, and 13 to rank 2.

%% file: 16proc.tex

\begin{figure}

\begin{tikzpicture}
	\draw[thick, scale=0.6] (0, 0) grid (4, 4);
	
	\node[anchor=center] (n0) at (0.3, 0.3+3*0.6) {P0};
	\node[anchor=center] (n1) at (0.3+0.6, 0.3+3*0.6) {P1};
	\node[anchor=center] (n2) at (0.3+2*0.6, 0.3+3*0.6) {P2};
	\node[anchor=center] (n3) at (0.3+3*0.6, 0.3+3*0.6) {P3};
	
	\node[anchor=center] (n4) at (0.3, 0.3+2*0.6) {P4};
	\node[anchor=center] (n5) at (0.3+0.6, 0.3+2*0.6) {P5};
	\node[anchor=center] (n6) at (0.3+2*0.6, 0.3+2*0.6) {P6};
	\node[anchor=center] (n7) at (0.3+3*0.6, 0.3+2*0.6) {P7};
	
	\node[anchor=center] (n8) at (0.3, 0.3+1*0.6) {P8};
	\node[anchor=center] (n9) at (0.3+1*0.6, 0.3+1*0.6) {P9};
	\node[anchor=center] (n10) at (0.3+2*0.6, 0.3+1*0.6) {P10};
	\node[anchor=center] (n11) at (0.3+3*0.6, 0.3+1*0.6) {P11};
	
	\node[anchor=center] (n12) at (0.3, 0.3) {P12};
	\node[anchor=center] (n13) at (0.3+0.6, 0.3) {P13};
	\node[anchor=center] (n14) at (0.3+2*0.6, 0.3) {P14};
	\node[anchor=center] (n15) at (0.3+3*0.6, 0.3) {P15};
	
	 \draw  (n0) to [thick, bend left=45]   (n1);
	 \draw  (n0) to [thick, bend right=45]   (n2);
	 \draw[red]  (n0) to [thick, loop above]   (n0);
	 
	  \draw  (n1) to [thick, bend right=45]   (n3);
	  \draw[red]  (n1) to [thick, bend right]   (n4);
	  \draw  (n2) to [thick, bend left=45]   (n3);
	  \draw[red]  (n2) to [thick, bend right]   (n8);
	   \draw[red]  (n3) to [thick, bend right]   (n12);
	  \draw  (n4) to [thick, bend left=45]   (n5);
	  \draw  (n4) to [thick, bend right=45]   (n6);	
	  \draw  (n5) to [thick, bend right=45]   (n7);	    
	  \draw  (n6) to [thick, bend left=45]   (n7);	
	  \draw[red]  (n5) to [thick, loop above]   (n5);
	  \draw[red]  (n6) to [thick, bend right]   (n9);
	  \draw[red]  (n7) to [thick, bend right]   (n13);
	  \draw  (n8) to [thick, bend left=45]   (n9);	    
	  \draw  (n8) to [thick, bend right=45]   (n10);	
	  \draw  (n9) to [thick, bend right=45]   (n11);	
	  \draw  (n10) to [thick, bend left=45]   (n11);	
	   \draw[red]  (n10) to [thick, loop above]   (n10);	  
	   \draw[red]  (n11) to [thick, bend right]   (n14);
	  \draw  (n12) to [thick, bend left=45]   (n13);	
	  \draw  (n12) to [thick, bend right=45]   (n14);
	  
	  \draw  (n13) to [thick, bend right=45]   (n15);
	  \draw  (n14) to [thick, bend left=45]   (n15);
	  \draw[red]  (n15) to [thick, loop above]   (n15);
	  
	  \path ($(n12) + (0,-1)$) to node[] {matrix columns}  ($(n15) + (0,-1)$);
	  \path ($(n12) + (-1,0)$) to node[rotate=90] {matrix rows}  ($(n0) + (-1,0)$);
 \end{tikzpicture} 
 \caption{Illustration of partitioning a matrix for NAS CG with 16 MPI processes; each process has 2 reduce partners (in black) and 1 transpose partner (red).}
 \label{fig:16procs}
\end{figure}

%% file: send-wrapper.tex
\section{The send wrapper}
\label{sec:send-wrapper}

The send wrapper needs to intercept each send call of the application; a receive equivalent is not required.
Unfortunately, the send wrapper is not simply overwriting the MPI calls, and the use of e.g. the MPI profiling interface is not possible.
The wrapper for each application needs to know the context and phase at which the sends were intercepted.

All of the send calls are intercepted by the wrapper, and the implementation is illustrated in Alg.~\ref{code:send-wrapper}.
First, a check is done if the send is into the future ($peer\_iters[me] < peer\_iters[dest] $). In this case, a send is cancelled, since the communication partner has already received the message.
This formulation generically applies to both restarted process, and some of the survivor processes, as illustrated in Fig. \ref{fig:failure-scenario}: 
\begin{itemize}
\item a restarted process in previous iterations in regard to survivors does not need to send them a message.
\item a surviving process in previous iterations in regard to other survivors, due to the inconsistencies caused by MPI's non-transactional semantics, does not need to send them a message.
\end{itemize}

If the send is not into the future, the send operation may be a normal send ($peer\_iters[me] == peer\_iters[dest]$) or a replay ($peer\_iters[me] > peer\_iters[dest]$).
For a normal send operation, we log the payload, unless we apply payload capping (discussed later in this section).
If the send is a replay, we read the stored payload log, and send it as a replayed MPI message to a communication peer (either a restarted process, or a survivor left behind during recovery).

\begin{algorithm}
\begin{lstlisting}[language=c,basicstyle=\small]
int send_wrapper(void * buf..., int current_iter) {
  // match send with the right iteration
  if (current_iter == peer_iters[dest]) {
    // regular send
    if (peer_iters[me] == peer_iters[dest])
       // sender-based payload logging
       // with potential capping capabilities
       append_log(buf, current_iter);
       MPI_Send(buf, ...);
     // replay message from buffer
    else if (peer_iters[me] > peer_iters[dest]) {
      log_buf = get_log(current_iter, ...);
      MPI_Send(log_buf, ...);
      ...
      }
      else {return 0;} // filter out sends into future
    }
    else {return 0;} // send irrelevant to sender
 }
\end{lstlisting}
\caption{Outline of send wrapper routine. The send wrapper replaces all MPI send calls of the application, and is also called by the replay routine.}
\label{code:send-wrapper}
\end{algorithm}

The high-level logic of the send wrapper is quite generic; unfortunately, the payload logging, which depend on the application data, is application-dependent in each of our extensions.

Because all payload logging is managed by the append\_log routine of Alg.~\ref{code:send-wrapper}, and within the application instead of the runtime, there are interesting optimisations which can be made to enable capping of logging message payloads.
The payloads of each application are conceptually not different from mini-checkpoints, in contrast to a runtime which only sees them as raw data.
These mini-checkpoints can be capped to save runtime and memory overheads; this, however, can limit the extent to which we can replay messages.
Yet it never affects global rollback -- we never compromise on checkpointing for global rollbacks.
The capping is easy to implement; details can be found in the source code.

%% file: capping.tex
\section{A Conditional Replay Routine for a Hybrid Rollback Protocol}
\label{sec:capping}

Upon failure detection, all processes synchronise to find everyone's current iteration, which represents the front line of computation.
Fig. \ref{fig:failure-scenario}, with survivors and restarted process in different iterations, represents such a front line.
The restarted process follows the normal application logic, having started from the last checkpoint iteration, with one notable exception -- all sends of the restarted process into the future are filtered out by the send wrapper of Alg.~\ref{code:send-wrapper}.

The recovery routine, which enables local rollback for all survivors, is at the core of our work, and illustrated in Alg.~\ref{code:replay-routine}.
The routine enables a hybrid protocol between local and global rollback, which is the task of survivor processes.

\begin{algorithm}
\begin{lstlisting}[language=c,basicstyle=\small]
void replay() {
  // gather the current iteration of each peer
  MPI_Allgather(peer_iters, ...);
  if (me == failed) { // restarted
    read_checkpoint();
  }
  else { // survivor
    maxit =  max(peer_iters);
    minit = min(peer_iters);
     // do we have the required payloads
     // stored for local rollback?
    bool local = (maxit % CP_INT < LOG_SIZE);
     //survivors don't roll back, but replay messages
    if (local) {
      for (int i=minit; i++; i<maxit) {
        // repicate send sequence of kernel
        // note: wrapper will use stored payloads
        send_wrapper(NULL, ..., i); 
        ...
      }
    }
    else { // global rollback necessary
      read_checkpoint();
    }
  }
} 
\end{lstlisting}
\caption{Outline of replay routine, called by all processes after a failure. All processes broadcast the front line of computation. If the required payloads are available, only the restarted process rolls back; otherwise all processes roll back.}
\label{code:replay-routine}
\end{algorithm}

If a log buffer is large enough to provide replay messages for the latest failure, a local rollback is enabled.
The local rollback is enabled as a sequence of replayed messages by all survivors.
Note that replayed messages are needed for restarted processes as well as survivors stuck in past iterations.
Since the purpose of a replay is for each process to send messages into the past, and since each process gets the global front line of computation at the start, no distinction between failed and survivng processes is required.
Once a process completes this process of supporting everyone left behind, it can safely resume with its normal iterations.
Note that the process supporting others may be in the past in relation to other processes after the replay routine, but it has successfully replayed all its messages at the end of the routine.

Otherwise, if the log buffer is not large enough to provide replay messages reaching into the most advanced current iteration, all survivors roll back together with the restarted process, resorting to global rollback.

\begin{figure}
\includegraphics[width=0.4\textwidth]{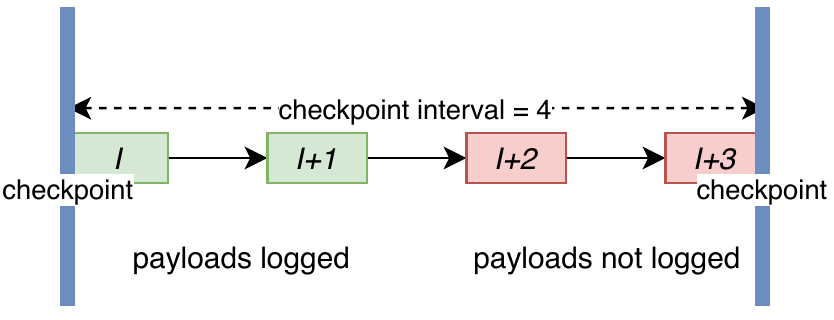}
\caption{Illustration of capping the payload buffer, and the resulting hybrid protocol between local and global rollback: failures where all survivors are within the green zone allow local rollback; failures where any survivors are in the red zone require global rollback.}
\label{fig:hybrid}
\end{figure}

This flexibility is illustrated in Fig. \ref{fig:hybrid}, which simply uses  a checkpoint interval of 4 (CP\_INT = 4), and a payload buffer limited to 2 iterations (LOG\_SIZE = 2).
The limited payload buffer reduced all memory and runtime overheads due to payloads to 50\% of what an efficient message logging protocol would require.
In our hybrid protocol, if after failure a survivor has progressed past these green iterations, it has stopped logging message payloads (red iterations).
Therefore, it cannot replay messages and local rollback is impossible; 
However, since we never compromise on global rollback, we can still perform global rollback.
If all survivors are within the first 2 green iterations when a failure is detected, local rollback is guaranteed.

We are not aware of any work in the MPI community implementing a hybrid mechanism between local and global rollback, which can gradually shift between either, depending on the user requirements.
While this modification is easy for message logging application kernels, whether this is possible with a message logging MPI runtime, acting on specific application hooks, is a more challenging research question.

The high-level logic of the replay routine is generic, but the sequence and arguments of the MPI sends is application-dependent.
It must accurately mimic the underlying application iterations.
Perhaps an advanced source-to-source compiler may be able to support this process in the future.

%% file: cg-outline.tex
\section{Extensions to CG and LULESH}
 \label{sec:loc}
 
 \subsection{CG from NAS Parallel Benchmarks}
 The NAS parallel benchmarks are documented by Bailey et al.~\cite{Bailey1991} and widely used in the HPC domain.
 We have chosen the NAS benchmark suite to demonstrate our approach mainly due to its popularity in the HPC community.
 We hope that by providing the extended version of the benchmark, researchers and developers will find it suitable for reproducible experiments and comparison.
 In terms of development, NAS benchmarks provide pre-defined correctness checks, which are extremely useful to us when integrating fault tolerance mechanisms.
 
 The focus on CG as iterative kernel is due to the fact that its communication pattern generates larger overheads than most other kernels, posing a significant challenge for message logging protocols.

%
%

The MPI version of the NAS CG benchmark is available in Fortran.
While the preferred option for us would be to only make incremental changes to the original Fortran code, various challenges have been reported when using ULFM with Fortran applications (see Weeks et al.~\cite{Weeks2018}).
Therefore, despite the effort, we decided to port the MPI version from Fortran to C.

\subsection{LULESH}
LULESH is a hydrodynamics code~\cite{Karlin2012} provided by LLNL, and a challenging mini-app for scientists to study in the HPC programming efforts towards exascale.
Our reference code is the provided checkpointing variant of the 1.0 LULESH code.

LULESH is interesting to us since it is a much larger, and more realistic HPC code, than the very compute-intensive but often benchmark-oriented CG kernel.
Despite the complexities of the code, and the significant larger number of communication calls across the code, we do not perceive LULESH as more challenging in its communication patterns than CG.
Its neighbourhood exchanges in 3D are reminiscent of stencil codes, which do not require collective communication.

\subsection{Problem Sizes}
For the NAS benchmarks, the problem sizes across classes A-E are well documented for CG; our tests are mostly limited to classes C-E.
For LULESH, we use the predefined $45^3$ elements per domain, and experiment with different process sizes, each with this fixed domain size (weak scaling).

\subsection{Summary of extensions}
Our extensions for either CG or LULESH consist of following steps:
\begin{itemize}
\item Introduce global rollback via traditional checkpointing, where unavailable (up to 220 lines for LULESH)
\item Introduce ULFM routines for revoking and recovering MPI communicators after failure.
This code ($\approx$ 200 lines) is a copy/paste from existing tutorial versions \cite{site:ulfm-tutorial}.
\item Introduce message logging, which consists of:
\begin{itemize}
\item buffer copying for transactional iterations, allowing transparent rollback within an iteration (10 lines for CG, 120 lines for LULESH)
\item modified sender calls, and replay routines (around 200 lines for both)
\end{itemize}
\end{itemize}

We summarise the lines of code (LOC) for each of these in Tab. \ref{tab:loc}.
\begin{table}
\begin{tabular}{|p{4cm}|p{1.75cm}|p{1,75cm}|}
-- &  Message logging CG & Message logging LULESH  \\
\hline
C/R & 50 & 220 \\
\hline
 ULFM-specific & 200 & 200 \\
 \hline
Message logging (replay, sender wrapper, payload logging)  & 200 & 200 \\
 \hline
 Temp buffers (for transactions) & 10 & 120 \\
 \hline
 Total & 1800 & 6500 \\
 \hline
\end{tabular}
\caption{LOC for message logging CG and message logging LULESH.}
\label{tab:loc}
\end{table}

The most lines specific to message logging are introduced for the replay routine of Fig.~\ref{code:replay-routine}, which mimics the communication pattern of each application kernel.
We have outlined our extensions previously; our message logging versions of CG and LULESH are open-sourced and freely available~\cite{message-logging-kernels}.

%% file: experiments.tex

\section{Experiments}
\label{sec:experiments}

\subsection{Verification}
Given the fact that we introduced message logging for the two kernels, the most significant outcome of our work is not performance, but the validation that the local rollback works correctly.
Our validation was as follows:
\begin{itemize}
\item For each checkpoint interval CP\_INT for either of the two kernels, crash a fixed MPI process in consecutive runs in iterations $i$, with $0 \leq i < CP\_INT$.
\item Use an application metric to compare the final result with the fault-free execution. NAS already provides invaluable verification tests ($\zeta$ and $r$ values). For LULESH, we read the final origin energy.
\end{itemize}

For either of the two kernels, and local rollback, we verified that errors in the replay routine, or not enabling transactional iterations, are among the main causes of incorrect results. 
Global rollback and failure-free runs produce correct results regardless of transactional iterations or replay routine.

\subsection{Performance measures}
Here, we give an overview of the runtime and memory overheads we measure for payload logging for either kernel.

\subsubsection{Experimental Platform}
We use the Cirrus cluster in EPCC (University of Ediburgh) for our experiments; each of 280 nodes consisting of 2 18-core Broadwell processors with 128 GB RAM per processor.
The interconnect is FDR infiniband.
Details are provided online~\cite{cirrus-online-hardware}.
We measured the bandwidth and latency across two nodes via NetPIPE; the maximum bandwidth is close to 50 Gbps, and the latency across nodes $\approx$ 2 microseconds.

We use ULFM 2.0 rc1 from November 2017 for all our experiments, configured with \verb1--with-ft1.

\subsubsection{Payloads -- memory overheads}

Here, we quantify the memory overheads for message payload logging. 
We can measure the per-log and the total heap space usage during an execution (we have modified static to dynamic memory allocation in the code) by using the popular Valgrind Massif tool for heap profiling.

\begin{table}
\begin{tabular}{|p{1.5cm}|p{1.5cm}|p{2.5cm}|p{1.5cm}|}
<kernel> -- <\#procs> & heap use without logging & payload size \textbf{per process and iteration} & percentage increase (heap) \\
\hline
cg.B.16 & $\approx 434$ MB & $\approx 450$ KB & 0.1\% \\
\hline
cg.C.16 & $\approx 1.08$ GB & $\approx 900$ KB & 0.09\% \\
\hline
cg.D.16 & $\approx 20$ GB & $\approx 9$ MB & 0.045\%  \\
\hline
LULESH.27 & $\approx 70$ MB & $\approx 500$ KB & 0.7\% \\
\end{tabular}
\caption{Memory overheads of payload logging for 16-process CG and 27-process LULESH.}
\label{tab:memory}
\end{table}

The payload overheads are given in Table \ref{tab:memory} for 16-proc CG runs and 27-proc LULESH runs.
Every process logs in the range of 450KB -- 9MB for each single iteration, which represents a memory overhead of up to 0.1\% for CG, and up to 0.7\% for LULESH, for each single iteration.

The accumulated overheads then depend on the checkpoint interval if we log all payloads, or if enabled, the capped payload logs of the hybrid protocol.
For example, if we checkpoint only once each 26 CG iterations, we would have payload overheads of up to 234 MB per MPI process for cg.D.16, which is over 1\% memory overhead per process.
If we checkpoint each 20 iterations for LULESH, we would have payload overheads of up to 20 MB per MPI process, which is $\approx$ 28\% memory overhead per process.
Payload capping can reduce these memory overheads.

\subsubsection{Payloads and temporary buffers -- runtime overheads}

\begin{table}
\begin{tabular}{|p{1.5cm}|p{1cm}|p{1cm}|p{1cm}|p{1cm}|p{1cm}|}
<kernel> -- <\#procs> & I/O & MPI time & payload copy & buffer copy &  total\\
\hline
lulesh.216 & 0.6 (2\%)  & 4.2 (12.7\%)   & 0.006 (0.02\%) & 2.7 (8\%) & 33.2 \\
\hline
cg.C.256 & 1.2 (21\%) & 1.4 (25\%) & 0.35 (6.2\%) & 0.053 (1\%) & 5.6 \\
\hline
 \end{tabular}
 \caption{Runtimes for a 75 iteration LULESH run (216 processes on 6 nodes), and a 75 iteration cg.C.256 run (256 procs on 8 nodes). Results are aggregations after manually instrumenting the phases of each kernel. Readings taken from rank 0; all units are seconds.}
 \label{tab:runtimes}
 \end{table}
 
For completeness, we show a more detailed breakdown of execution time in Table \ref{tab:runtimes}.
It includes I/O overheads, time spent in MPI calls, payload copy overhead, and buffer copy overhead to enable transactional iterations; the wall clock times for LULESH and CG are also included, and used to calculate the relative contributions of the corresponding phases.

For all measured times, we have manually instrumented both CG and LULESH across different phases.
We also show the percentage of these times to the overall wall clock times.
Since CG works with larger data volumes than LULESH, the MPI and I/O times represent a higher percentage of the overall runtime.
We are more interested in the overheads for payloads, and also for the temporary buffers enabling transactional iterations, both parts of our extensions.
CG incurs significant payload overheads of up to 6.2\% with payload logging, compared to the almost non-existent payload overheads for LULESH; again, this relates to the significant communication exchange for CG.
These results are consistent with the results of related work with message logging protocols.
In contrast, the buffer duplication to enable transactions comes at a high price for LULESH (8\%), compared to the low overhead for CG (1\%).
Since we choose to create deep copies of the entire LULESH domain each iteration, which represents all application data, this is also not surprising. 
In contrast, for CG we only copy a few persistent vectors each iteration.

Again, payload capping can reduce the runtime overheads, which are significant for CG.
The copy into temporary buffers cannot be reduced in the same way, since transactional iterations are a requirement for the proposed approach; however, since we copy the entire application data for LULESH, various optimisations seem feasible.

\subsubsection{Wall clock times}

In this section, we measure what are the consequences of local and global rollback to recompute, as well as overall wall clock times.

\begin{figure}
\includegraphics[width=0.5\textwidth]{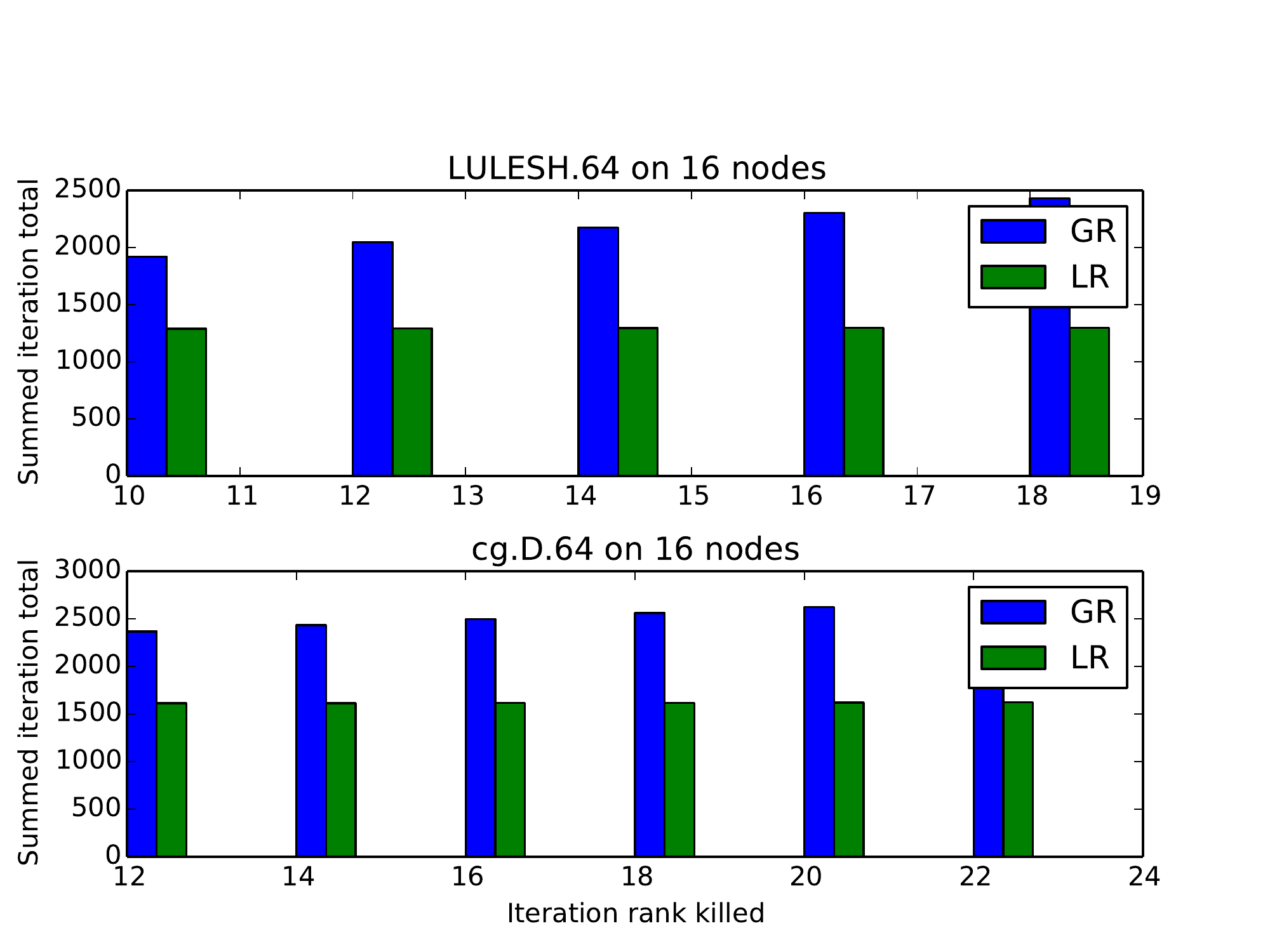}
\caption{Total summed iterations for 16-node runs with cg.D.64 and LULESH.64, depending on where a failure is triggered, and type of rollback.}
\label{fig:total-it}
\end{figure}

\begin{figure}
\includegraphics[width=0.5\textwidth]{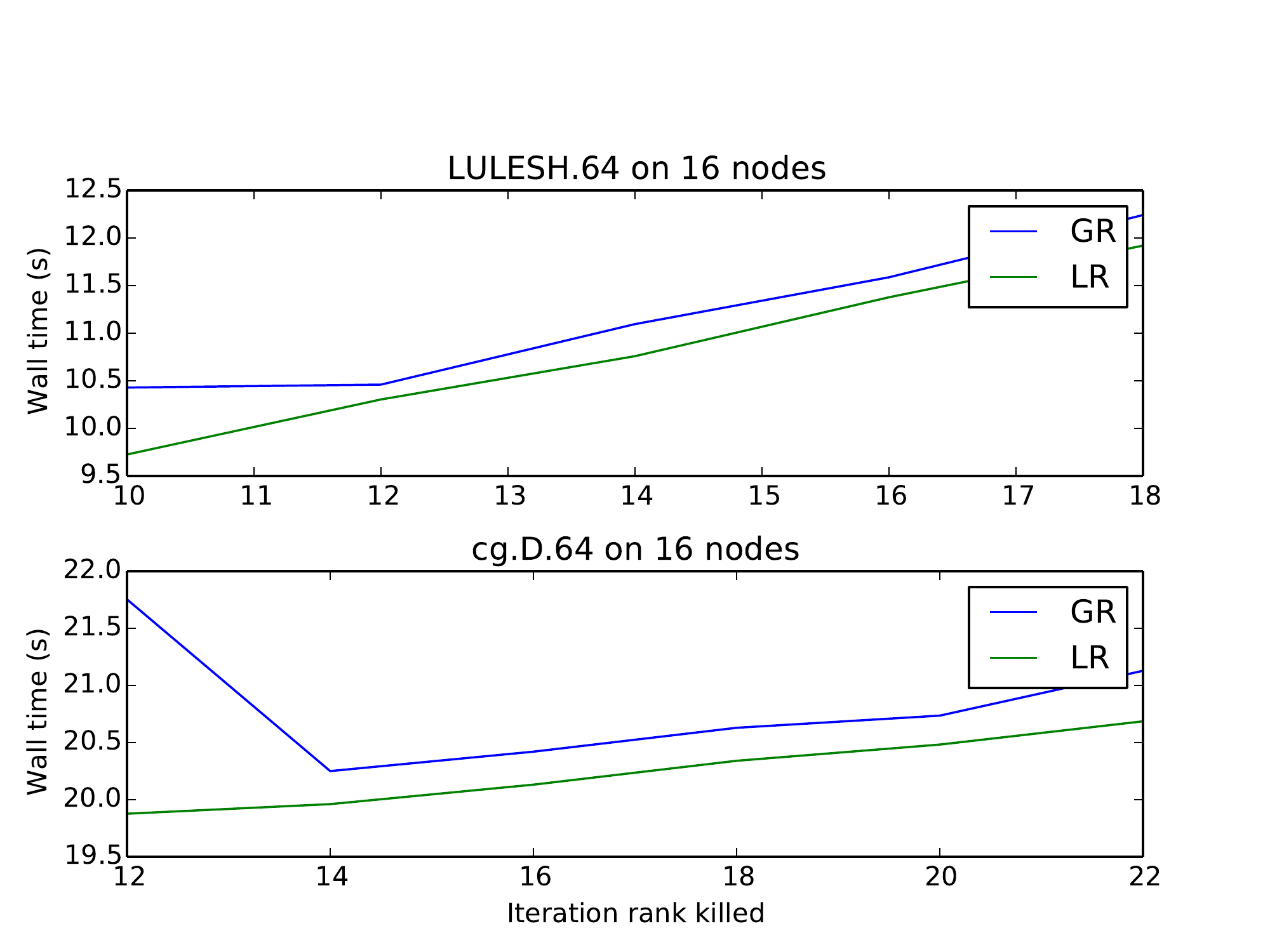}
\caption{Wall clock time for 16-node runs with cg.D.64 and LULESH.64, depending on where a failure is triggered, and type of rollback. The plot indicates that local rollback does not offer significant time reduction in overall execution time, compared to global rollback.}
\label{fig:walltime}
\end{figure}

We outlined our non-trivial setting earlier in this section --  we crash a fixed MPI process in consecutive runs for iterations $i$, with $0 \leq i < CP\_INT$. 
We set the CP\_INT for CG to 25 iterations, and for LULESH to 20 iterations.
Within this setting, we first determine the total number of compute iterations, summed across all processes, as a function of when a crash happens (see Fig.~\ref{fig:total-it}).
For the same runs, we measure the wall clock time to completion after failure for all runs; the times are shown in Fig.~\ref{fig:walltime}.
As we can see, local rollback as expected decreases consistently the number of recomputed iterations, since only restarted processes roll back.
However, this does not lead to significantly faster overall execution times.
Since the restarted process is on the critical path of recovery both for local and global rollback, the results are not surprising.
We believe in general faster execution time should not be expected of local rollback.
Rather, local rollback needs to be combined with other techniques to yield immediate end-to-end benefits; we have previously demonstrated such techniques~\cite{Dichev2018} by applying frequency scaling to the idle processes.
Another potential optimisation is to collocate with other processes during recovery on idle processors.

\begin{figure}
\includegraphics[width=0.5\textwidth]{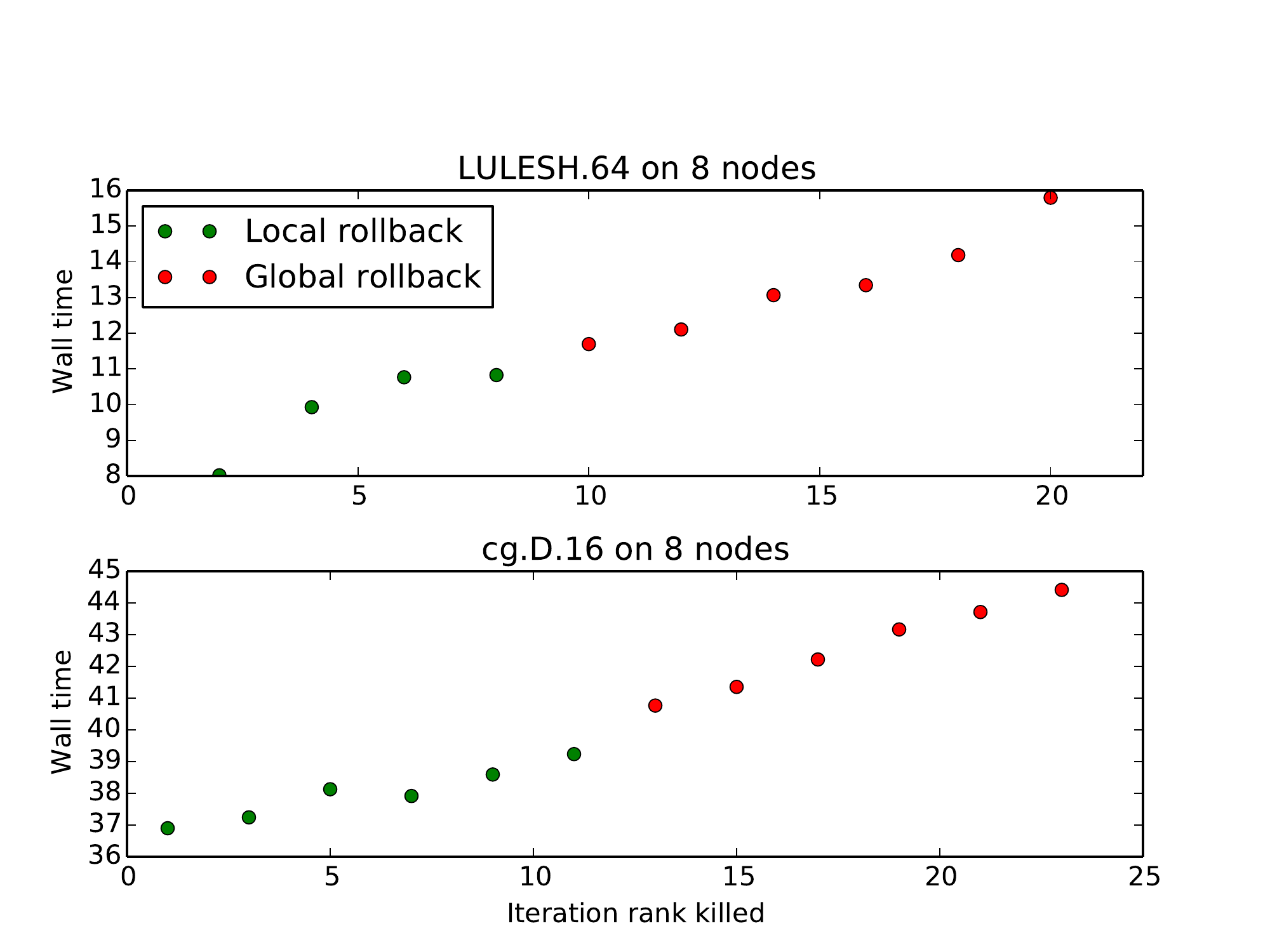}
\caption{Demonstration of the hybrid protocol which automatically switch between local and global rollback. In this example, the payload capacity is $\frac{CP\_INT}{2}$ for each kernel.}
\label{fig:hybrid2}
\end{figure}

Finally, we have validated the hybrid local-global rollback, by setting the capacity of the payload logs to $LOG\_SIZE = \frac{CP\_INT}{2}$.
As expected, the protocol gracefully switches between local rollback, if a failure strikes early during the $CP\_INT$ window, and global rollback, if the failure strikes later.
We show the runtime results for the hybrid approach, again testing different iterations of failure for a fixed process, in Fig.~\ref{fig:hybrid2}.
While the hybrid protocol dynamically switches between local and global rollback, always providing correct results, the overall wall clock time linearly increases either way, since when a failure strikes later, more iterations need to be repeated by one process, or all processes.

%% file: conclusion.tex

\section{Conclusion}
\label{sec:conclusion}
In this work, we introduced message logging capabilities into two popular HPC kernels, the NAS CG benchmark, and the mini-app LULESH.
One of our main motivators for doing so is the fact that message logging MPI runtimes get published today, but are not easy to deploy by anyone but a few experts in these protocols.
We introduced a methodology for implementing message logging into HPC kernels; this included using additional temporary buffers to enable what we called transactional MPI iterations, so that the rollback of a process does not break its data consistency.
We then introduced sender-based payload logging and a replay mechanism, which enabled us to restore the inter-process data consistency.
All of the above techniques were presented in a generic way, but aspects such as the payload logging or sequence of replayed messages remained kernel-specific.
As a performance optimisation which has never been proposed before, we were also able to introduce a hybrid rollback protocol, which gracefully shifts from local to global rollback, if we choose to decrease the available memory for payload logging.

While our focus is on validation rather than performance, we presented experimental results, which show that local rollback is fully functional, and therefore a complete replacement for message logging MPI runtimes, for these kernels, and carries less overheads than most implementations (e.g. no event logger required).
However, the idling processes due to reduced rollback did not immediately reduce overall runtime; we have pointed to techniques which exploit the reduced rollback to bring end-to-end runtime or energy benefits.
We hope our methodology and codebase may be useful to the community for experimenting with message logging protocols as compact HPC kernels extensions, as a replacement of complex and difficult to set up message logging runtimes.

\section*{Acknowledgment}

This work used the Cirrus UK National Tier-2 HPC Service at EPCC (http://www.cirrus.ac.uk) funded by the University of Edinburgh and EPSRC (EP/P020267/1).